\documentclass[aip,apl,amsmath,amssymb,reprint]{revtex4-1}

\usepackage{graphicx}
\usepackage{dcolumn}
\usepackage{bm}

\begin{document}

\title{Luminescence decay dynamics of germanium nanocrystals in silicon}

\author{B. Julsgaard}
\altaffiliation{Electronic mail: brianj@phys.au.dk.}
\author{P. Balling}
\author{J. Lundsgaard Hansen}
\author{A. Svane}
\author{A. Nylandsted Larsen}
\affiliation{Department of Physics and Astronomy, Aarhus University,
  DK-8000 Aarhus C, Denmark}

\date{\today}
             
\begin{abstract}
  The dynamics of the luminescence decay from germanium nanocrystals
  embedded in crystalline silicon has been studied for temperatures
  varied between 16 K and room temperature. At room temperature the
  characteristic decay time is of the order of 50 nanoseconds while it
  extends into the microsecond range at low temperatures. The decay is
  dominated by non-radiative processes, which show a typical thermal
  activation energy of a few meV.
\end{abstract}

\maketitle

Luminescence from germanium (Ge) nanocrystals embedded in silicon (Si)
is interesting since the emission wavelength can be tuned in the range
around $1.5\:{\mu}\mathrm{m}$ used by standard fiber communication
technology. Hence such nanocrystals present one of several routes
toward integration of optical and electronic functionality on a
silicon chip. Previously, the spectral dependence of the luminescence
from Ge (or SiGe) nanocrystals has been studied as a function of
various growth parameters, \cite{Sunamura.JCrystGrowth.157.265(1995),
  Bremond.MicroelectronicsJ.30.357(1999),
  Schmidt.ApplPhysLett.75.1905(1999),
  Schmidt.ApplPhysLett.77.2509(2000),
  Dashiell.ApplPhysLett.79.2261(2001),
  Larsson.PhysRevB.73.195319(2006),
  Tonkikh.PhysStatSolRRL.4.224(2010)} and experimental parameters
\cite{Fukatsu.ApplPhysLett.71.258(1997),
  Larsson.PhysRevB.73.195319(2006),
  Adnane.ApplPhysLett.96.181107(2010)} such as sample temperature,
laser excitation power and wavelength, and stress, often in search of
a deeper understanding of the Ge/Si system (e.g.~band edge alignment
\cite{ElKurdi.PhysRevB.73.195327(2006)}). Some of these studies have
demonstrated room-temperature luminescence,
\cite{Sunamura.JCrystGrowth.157.265(1995),
  Bremond.MicroelectronicsJ.30.357(1999),
  Schmidt.ApplPhysLett.77.2509(2000),
  Tonkikh.PhysStatSolRRL.4.224(2010)} but luminescence life time
measurements have only been presented to a very limited
extent.\cite{Fukatsu.ApplPhysLett.71.258(1997),
  Bremond.MicroelectronicsJ.30.357(1999)}

In this letter we study the dynamics and spectral characteristics of
the luminescence from Ge nanocrystals as a function of temperature
from room temperature down to 16 K. By lowering the temperature, a
significant increase in the decay time is observed as the most
pronounced effect, leading to our conclusion that the decay is
dominated by non-radiative processes. In addition, the decay behavior
can be interpreted consistently with models of Ge/Si nanocrystals
published previously. \cite{Larsson.PhysRevB.73.195319(2006),
  ElKurdi.PhysRevB.73.195327(2006)}

\begin{figure}[b]
\includegraphics[width=\linewidth]{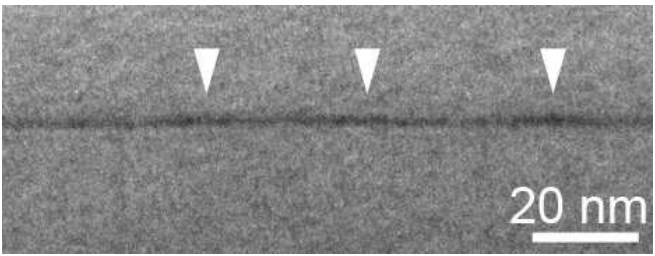}
\caption{\label{fig:TEM}Cross-sectional transmission electron
  micrograph showing that the Ge nanocrystals are barely rising above
  the wetting layer and the lateral size is of the order of 20 nm.}
\end{figure}

Self-assembled Stranski-Krastanow Ge nanocrystals were prepared by
molecular beam epitaxy. On a Si(100) wafer ($\approx
100\:\Omega\mathrm{cm}$) a 200 nm Si buffer layer was grown followed
by a layer of Ge, which was deposited without rotating the wafer
leading to a non-uniform Ge layer thickness. Finally, a 50 nm Si
capping layer was added. A surfactant
\cite{Tonkikh.PhysStatSolRRL.4.224(2010)} of approximately 0.01
mono-layer of antimony (Sb) was used, and the growth temperature was
$530\:^{\circ}\mathrm{C}$. Five pieces with nominal Ge layer
thicknesses of 8.25, 9.00, 9.75, 10.50, and 11.50 {\AA} have been cut
from the wafer. A transmission electron micrograph (TEM) obtained for
the sample with nominal Ge-layer thickness of 9.75 {\AA} shows that
nanocrystals have been formed with diameters around 20 nm, see
Fig.~\ref{fig:TEM}.

The samples were mounted in a closed-cycle helium cryostat allowing
for sample temperatures down to 16 K. An amplified 1 kHz femto-second
Ti:sapphire laser system was frequency doubled in order to produce 100
fs pulses at 400 nm (3.1 eV) for exciting the Ge nanocrystals. The
laser pulse energy entering the sample was typically $6\:{\mu}J$,
focused by a cylindrical lens with 100 mm focal length leading to a
fluence of the order of $6\cdot 10^{-4}\:\mathrm{J/cm^2}$. The
nanocrystal fluorescence was collected into a McPherson 218
spectrometer equipped with a liquid-nitrogen-cooled photo-multiplier
tube (Hamamatsu R5509-73).

\begin{figure}
\includegraphics{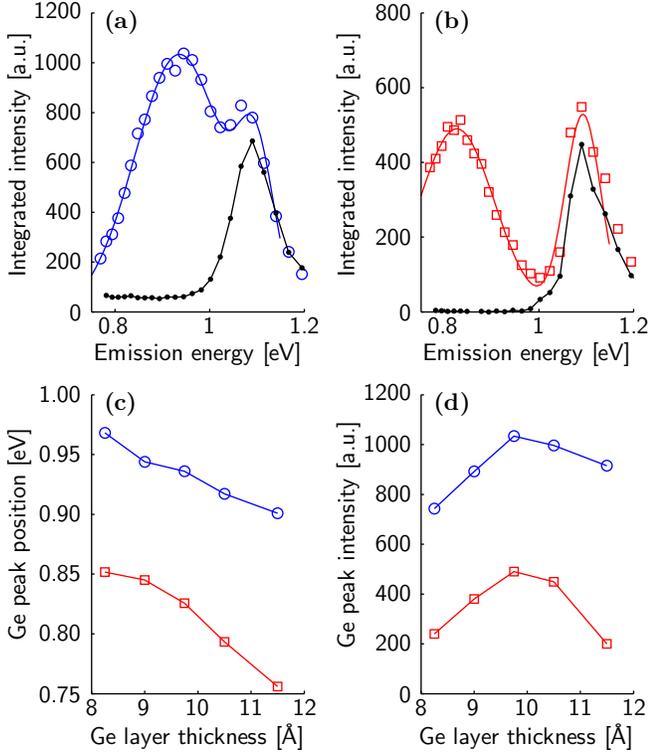}
\caption{\label{fig:RoomTempTRES}(Color on-line) Panels a and b show
  time-resolved emission spectra of the sample with nominal Ge
  thickness of 9.75 {\AA} at 294 K. The time range of the spectra are
  $0 \le t \le 5\:\mathrm{ns}$ (blue circles) and $25\:\mathrm{ns} \le
  t \le 200\:\mathrm{ns}$ (red squares) for panels a and b,
  respectively. The data can be modeled approximately by a sum of two
  Gaussian peaks (solid lines). Spectra from a sample without Ge are
  shown for comparison (black dots). Panels c and d show (with symbols
  corresponding to those of panels a and b) the fitted position and
  height, respectively, of the lower-energy peak in the two-Gaussian
  model for various Ge layer thicknesses.}
\end{figure}

Time-resolved emission spectra obtained at room temperature are shown
in Fig.~\ref{fig:RoomTempTRES}(a,b). The spectra can be fitted
reasonably to a sum of two Gaussian functions allowing an easy
determination of the general trend of the emission
characteristics. The peak just below 1.1 eV arises from band-edge
recombination in bulk silicon. The lower-energy peak is only present
in samples containing Ge, and its position changes toward lower
energies during the first few tens of ns (compare panels a and b in
Fig.~\ref{fig:RoomTempTRES}). Decreasing the Ge-layer thickness leads
to an increase of the emission energy for both the early-time and
late-time peaks (Fig.~\ref{fig:RoomTempTRES}c). This blue-shift in
emission energy together with the requirement that Ge must be present
indicates that the light is emitted from quantum confined Ge-related
structures. Although emission from the Ge wetting layer (WL) is
usually not observed in steady-state photo-luminescence experiments
for samples grown at $530\:^{\circ}\mathrm{C}$, at least the emission
energy of the early-time peak is in the correct
range\cite{Larsson.PhysRevB.73.195319(2006)}. We attribute the
late-time peak to emission from Ge nanocrystal islands in agreement
with previous observations. \cite{Sunamura.JCrystGrowth.157.265(1995),
  Bremond.MicroelectronicsJ.30.357(1999),
  Schmidt.ApplPhysLett.75.1905(1999),
  Schmidt.ApplPhysLett.77.2509(2000),
  Dashiell.ApplPhysLett.79.2261(2001),
  Larsson.PhysRevB.73.195319(2006),
  Tonkikh.PhysStatSolRRL.4.224(2010),
  Fukatsu.ApplPhysLett.71.258(1997)} As can be seen in
Fig.~\ref{fig:RoomTempTRES}d, the sample with a nominal Ge-layer
thickness of 9.75 {\AA} emits most light and we selected this sample
for further investigations at varying temperature.

\begin{figure}
\includegraphics{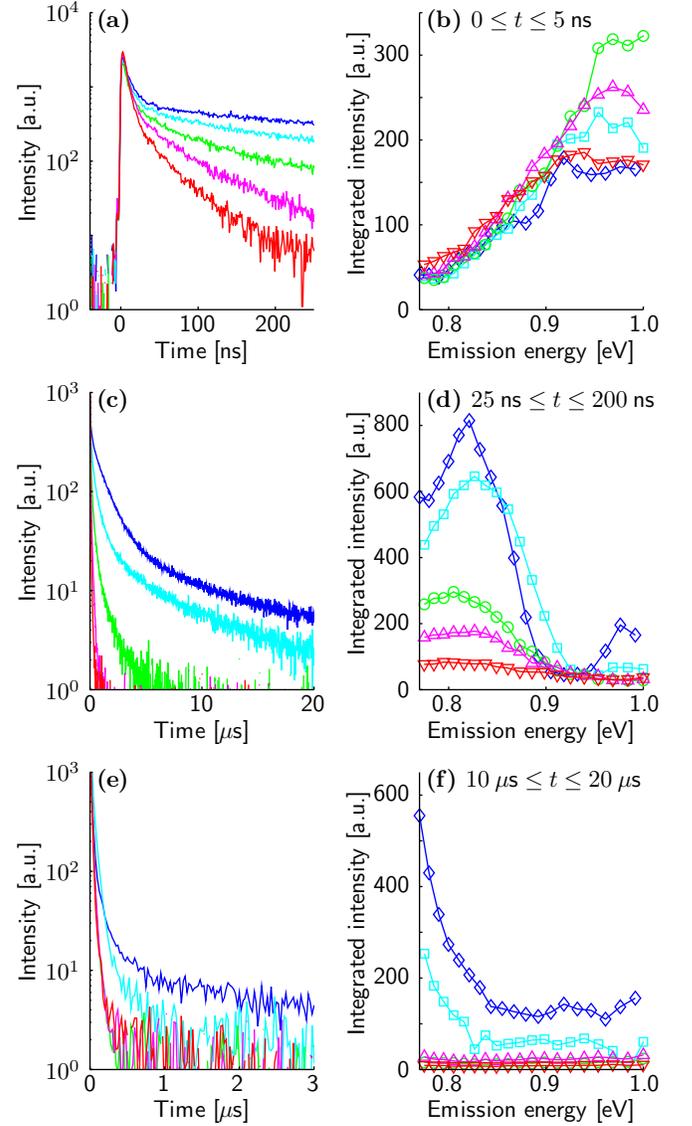}
\caption{\label{fig:DataComparison}(Color on-line) Decay curves
  (panels a, c, e) and time-resolved emission spectra (panels b, d, f)
  plotted for various temperatures: 294 K (red tip-down triangles, 200
  K (purple tip-up triangles), 100 K (green circles), 40 K (cyan
  squares), and 22 K (blue diamonds). In panels (a, c, e) the upper
  graph shows the 22 K data, while the temperature increases through
  the above-mentioned values for the subsequent lower curves. The
  emission energy is 0.775 eV for the curves in panels a and c, and
  0.886 eV in panel e. Time ranges for the spectra in panels b, d, and
  f are mentioned on the graphs.}
\end{figure}

The separation of time scales into an early-time part ($\approx$10 ns
time scale) and a late-time part is evident when considering the
luminescence decay curves shown in
Fig.~\ref{fig:DataComparison}a. This graph shows decay curves obtained
at the emission energy of 0.775 eV for varying temperatures. The
initial peak is essentially independent on temperature, while the
longer-time decay changes significantly with temperature. To clarify
this effect further, we obtained time-resolved emission spectra in the
time range, $0 \le t \le 5\:\mathrm{ns}$, which selects primarily the
initial peak. The results have been plotted in
Fig.~\ref{fig:DataComparison}b, and the similarity for varying
temperature is evident. This challenges the interpretation that this
light originates from WL emission: The insensitivity to temperature -
i.e.~the fact that the spectral feature remains broad - must be
clarified in a more systematic study before a clear identification can
be made. On the contrary, when setting the time range to
$25\:\mathrm{ns} \le t \le 200\:\mathrm{ns}$, and thereby selecting
the late-time part of Fig.~\ref{fig:DataComparison}a, the emission
characteristics of Ge islands are recovered as shown in
Fig.~\ref{fig:DataComparison}d. The decrease in intensity with
increasing temperature corresponds to observations reported
previously. \cite{Larsson.PhysRevB.73.195319(2006)}

In order to investigate further the Ge island emission, we plot decay
curves for longer time scales in Fig.~\ref{fig:DataComparison}c (0.775
eV) and Fig.~\ref{fig:DataComparison}e (0.886 eV). These curves show
several features: (i) The decay time increases significantly when
lowering the temperature, (ii) the low-energy side of Ge island
emission decays significantly slower than the high energy side, and
(iii) there is an asymptotic limit for the decay curves; zooming in on
the high-temperature curves in panels c and e of
Fig.~\ref{fig:DataComparison} would reveal essentially the same decay
behavior. From these observations we suggest that two distinct decay
mechanisms co-exist, where one is much more robust to high
temperatures than the other.

We also note that for the lowest temperatures a distinct decay
component with a very slow decay time arises. If the uppermost graph
in Fig.~\ref{fig:DataComparison}c is fitted to a single-exponential
function in the range, $10\:{\mu}s \le t \le 20\:{\mu}s$, a decay time
of $13\:{\mu}s$ is found, while in the range, $1\:{\mu}s \le t \le
3\:{\mu}s$, the same procedure yields $1.6\:{\mu}s$. In order to
reveal the spectral behavior of the very-long-time component, a
time-resolved emission spectrum for the time range, $10\:{\mu}s \le t
\le 20\:{\mu}s$, is shown in Fig.~\ref{fig:DataComparison}f. We see
that there is both a flat background above 0.85 eV and an increased
intensity at lower energies. The latter suggests that the
very-long-time emission originates at least partly from the Ge islands
with a down-shifted emission energy (by 50 meV or more, compare
Figs.~\ref{fig:DataComparison}d and f). Assigning the
$\ge$10-$\mu\mathrm{s}$ component to spatially indirect transitions
across the Ge/Si boundary and the $\approx$1-$\mu\mathrm{s}$ component
to spatially direct transitions within the Ge nanocrystal would be
consistent with previous
results. \cite{Larsson.PhysRevB.73.195319(2006)}

\begin{figure}
\includegraphics{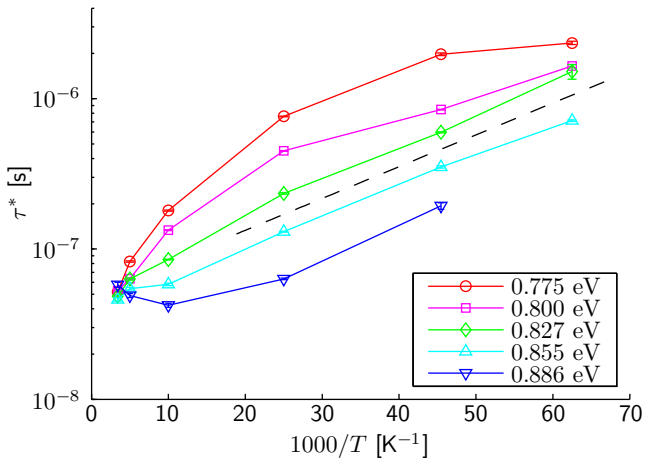}
\caption{\label{fig:TauStar}(Color on-line) The characteristic decay
  time, $\tau^*$, as a function of reciprocal temperature plotted for
  various emission energies. The error bars denote the statistical
  uncertainty from the fitting procedure. The dashed line corresponds
  to an activation energy of $4.2\:\mathrm{meV}$.}
\end{figure}

In order to make a general comparison of the decay dynamics for
various temperatures and emission energies, we make a
multi-exponential fit to each decay curve modeled as: $f(t) = d_0 +
\sum_{j=1}^N A_j \exp(-t/\tau_j)$. Here $d_0$ is the dark count level,
and the number of terms, $N$, is adjusted in order to fit the entire
decay curve reasonably (in practice $N$ varies between 2 and 4). The
fastest term will model the initial peak in
Fig.~\ref{fig:DataComparison}a, which has a spectral characteristic
(Fig.~\ref{fig:DataComparison}b) different from that of Ge islands. In
practice the fastest timescale varies between 6 ns and 12 ns within
the temperatures and emission energies studied here. In the following
analysis we exclude this fastest term in order to model more closely
the actual Ge island emission. We define the characteristic decay
time: $\tau^* = \text{Area}/\text{Amplitude} = (\sum_{j=2}^N
A_j\tau_j)/(\sum_{j=2}^N A_j)$, where the ``Area'' (under the decay
curve) is the integrated intensity and ``Amplitude'' is the initial
intensity at zero time. Since $\tau^*$ is proportional to the
integrated intensity, it represents in a direct way how an increased
decay time leads to a more efficient light emission in total. In
Fig.~\ref{fig:TauStar} we show Arrhenius plots of $\tau^*$ as a
function of $1000/T$ for different emission energies. Most pronounced,
when lowering the temperature the decay time becomes slower. The fact
that $\tau^*$ increases significantly with decreasing temperature
while the initial decay amplitude varies only modestly (extrapolating
the late-time-part of the decay curves in
Fig.~\ref{fig:DataComparison}a back to zero time will lead to roughly
the same initial value) proves that the decay is dominated by
non-radiative processes. Although the upper-most curve in
Fig.~\ref{fig:TauStar} shows some effects of saturation, the general
trend is a continued increase of $\tau^*$ for decreasing temperature
and hence that non-radiative processes are dominant also at 16 K. We
conclude that the radiative decay time must be at least well into the
microsecond range. Such long radiative decay times are characteristic
for indirect-band-gap
materials.\cite{Hybertsen.PhysRevLett.72.1514(1994)} We also note in
Fig.~\ref{fig:TauStar} that the non-radiative decay processes are
activated on the few-meV energy scale with little dependence on the
emission energy (compare the data curves with the dashed line in
Fig.~\ref{fig:TauStar}). Assigning an energy barrier of this magnitude
to a metastable electronic state within the Ge nanocrystal is
consistent with the calculations of
Ref. \cite{ElKurdi.PhysRevB.73.195327(2006)} The lower bound around 50
ns is characteristic for the asymptotic decay curves discussed in
relation to Fig.~\ref{fig:DataComparison}(c,e) and may arise, as
discussed, from a separate emission mechanism.

In conclusion, we have measured the characteristic decay times for Ge
nanocrystals as a function of emission energy and temperature. We
demonstrate that the decay is dominated by non-radiative recombination
mechanisms. A shallow few-meV energy barrier of electronic states
within Ge nanocrystals is suggested as responsible for the
non-radiative decay process.

This work was supported by The Danish Council for Independent Research
$\vline$ Natural Sciences and by the Carlsberg Foundation. We thank
J. Chevallier for assistance with TEM measurements.

%

\end{document}